\documentclass[sigconf, 9pt,  screen]{acmart}
\startPage{1}
\renewcommand\footnotetextcopyrightpermission[1]{}

\setcopyright{none}

\usepackage{amsmath,amsfonts}
\usepackage{algorithmic}
\usepackage{graphicx}
\usepackage{textcomp}
\usepackage{xcolor}
\usepackage{listings}
\usepackage{tabularx,booktabs}
\usepackage{float}
\usepackage{subcaption}

\usepackage{url} 

\definecolor{darkgreen}{rgb}{0.0, 0.4, 0.0}
\lstdefinelanguage{MLIR}
{
    alsoletter={.},
    keywords=[1]{
      add,addf,affine,affine.apply,affine,
      alloc,br,cmpi,constant,delay,dim,dma_start,dma_wait,
      for,func,affine.load,mulf,affine.store,load,store, 
      reduce,reshape,matmul,reduce_sum,mhlo,while,Conv2D,xla_hlo,xla_lhlo,return,compare,iota,dynamic_slice,dyname_update_slice,Add,
      arith.mulf,arith.addf,func.call,affine.for,func.func,
      hir.load, hir.store, hir.func, hir.for,hir.if,hir.alloca,private,
    hw.constant, hir.send, hir.recv,hir.memref.extract, hir.yield, hir.delay, by,
  hir.call,hir.next_iter,hir.return, hir.func.extern, at, step, to, from, port, ports, bank, comb.add,comb.extract, arith.constant},
    keywordstyle=[1]\color{magenta}\ttfamily\bfseries,
    keywords=[2]{
      int,memref,tensor,vector,i1,i32,f32,f64,i64,i6,tuple,const,hir.memref,hir.const,hir.time,hir.group,hir.array.hir.func
    },
    keywordstyle=[2]\color{blue}\bfseries,
    keywords=[3]{
      func, with, affine_map, affine_set, def, dense, step, tstep, iter_time,
      time, at
    },
    keywordstyle=[3]\color{magenta}\bfseries,
    morecomment=[l]{//},
    commentstyle=\color{darkgreen},
    backgroundcolor=\color{white},
    mathescape,
    tabsize=2,
    basicstyle=\small\ttfamily
}

\begin{document}


\title{Automatic multi-dimensional pipelining for high-level synthesis of dataflow accelerators}
                                        
 \author{Kingshuk Majumder}
\affiliation{
  \department{Computer Science and Automation}  
  \institution{Indian Institute of Science}     
  \city{Bangalore}
  \country{India}
}
\email{kingshukm@iisc.ac.in}

\author{Uday Bondhugula}
\affiliation{
  \department{Computer Science and Automation}  
  \institution{Indian Institute of Science}     
  \city{Bangalore}
  \country{India}
}
\email{udayb@iisc.ac.in}
                                       
\begin{abstract}
In recent years, there has been a surging demand for edge computing of image
processing and machine learning workloads. This has reignited interest in the
development of custom hardware accelerators that can deliver enhanced
performance and improved energy efficiency. These workloads frequently
demonstrate affine memory accesses and constant loop bounds. In this paper, we
introduce an ILP-based automatic scheduler for high-level synthesis, with a
specific emphasis on aggressive pipelining to enhance parallelism.

Applications may present opportunities for pipelining along various dimensions,
such as pipelining iterations within individual loops or pipelining across
producer-consumer loops. Often, these distinct pipelining strategies can be
combined to maximize parallelism. In this study, we propose a unified Integer
Linear Programming (ILP) formulation that can identify pipelining opportunities
along multiple loop and scalar dimensions. Our multi-dimensional pipelining
technique encompasses both inner loop pipelining and dataflow optimizations of
Vitis HLS, while also being capable of handling more general memory access
patterns compared to the dataflow optimization in Vitis HLS. Furthermore, our
approach enables the generation of statically scheduled circuits, leading to
improved resource efficiency.

We have integrated our scheduler into a high-level synthesis compiler framework
(HIR) based on MLIR and conducted performance evaluations. Our findings reveal
that our scheduler, in comparison to Vitis HLS, can achieve more aggressive
pipelining across multiple producer-consumer loop nests, resulting in reduced
overall execution latency. The producer-consumer pipelined execution facilitated
by our scheduler yields an average performance improvement of 2.42X across a
set of representative benchmarks when compared to only loop pipelining.
Furthermore, we achieved an average performance improvement of 1.30X over Vitis
HLS with dataflow optimizations.
\end{abstract}


\maketitle
\pagestyle{plain}

\section{Introduction}
The demand for improved image processing and advanced machine learning models on
edge devices has outpaced the computing capabilities of traditional CPUs and
GPUs. As a result, there is a renewed demand for domain-specific accelerators
that offer higher power efficiency and improved performance. Tasks such as video
compression, which requires significant computational power and performance, are
already being offloaded to specialized accelerators in modern CPUs. While
general-purpose GPUs have historically offered high parallelism for task
parallel workloads, they have also incorporated specialized accelerators in
their design to keep up with the increasing complexity of machine learning
models in recent years.

FPGAs provide a cost-effective platform for deploying domain-specific
accelerators, thanks to their ability to customize accelerator design, memory
hierarchy tailored to the application, and software-managed caching. FPGA-based
accelerators strike a good balance between power efficiency and performance.
In recent years, high-level synthesis (HLS) has become a practical and effective
alternative to Verilog-based hardware design for FPGAs. HLS compilers can
automatically pipeline loops, perform resource allocation and reuse, and insert
pipeline registers to achieve timing closure.  This allows hardware designers to
concentrate on the high-level architecture of their design while leaving the
low-level optimizations, parallelization, and pipelining tasks to the HLS
compiler.
Many domain-specific and general-purpose
tools~\cite{auerbach12dac,hegarty2014darkroom,hegarty2016rigel,reiche14codes,
chugh16pact,vivado,hdl-coder,dase06ieeecs,najjar03computer,kingshuk21hir,asplos21calyx}
have been built in the past few decades. 
Bacon et al.~\cite{rabbah13acm} did a comprehensive survey of various HLS techniques.

A prevalent approach to improve the quality of hardware generated in high-level
synthesis is to identify parallelizable patterns that are commonly found
in various workloads and then develop specialized optimizations specifically
tailored to those patterns of parallelism. 
PolySA~\cite{cong2018polysa} and AutoSA~\cite{wang2021autosa}
apply specialized optimizations to generate efficient hardware for systolic
arrays. 
Similarly, many prior HLS frameworks  
~\cite{iccad18soda,cgo21stencilflow,tian2023sasa,dac2014nupart} have
focused on optimizing high-level synthesis for stencil computations.

Image processing and machine learning workloads involve multiple window-based
and point-wise operations, each of which is represented using nested loops
with affine memory accesses. While pipelining inner loops is a common practice
in modern commercial compilers, more sophisticated pipelining techniques, such
as pipelining among the producer and consumer operations, are not
well-supported. For example, dataflow optimizations in Vitis HLS impose strict
constraints on the data access patterns of the producer and consumer loop nests
and do not support scenarios with multiple producers or multiple consumers.
As a result, the programmer has to introduce extra tensor copying steps to
convert a design with multiple consumers to a single-producer-single-consumer
(SPSC) design. This transformation adds extra intermediate buffers in the design
to keep the copies.

\begin{figure*}
  \centering
  \begin{subfigure}{0.4\linewidth}
  \centering
\begin{lstlisting}[xleftmargin=5pt,numbers=left,language=C++,breaklines=true,frame=bt,basicstyle=\scriptsize]
//Producer loop-nest.
for (int i = 0; i < 3; i++) {
  for (int j = 0; j < 3; j++)
#pragma pipeline
    for (int u = 0; u < 2; u++)
#pragma unroll
      for (int v = 0; v < 2; v++)
#pragma unroll
        convX[i][j] += image[i+u][j+v] * wx[u][v];
}

//Consumer loop-nest.
for (int i = 0; i < 3; i++) {
  for (int j = 0; j < 3; j++)
#pragma pipeline
     for (int u = 0; u < 2; u++)
#pragma unroll
       for (int v = 0; v < 2; v++)
#pragma unroll
         convY[i][j] += convX[i+u][j+v] * wy[u][v];
}
\end{lstlisting}
  \caption{Two convolution operations in a series.}
  \label{fig:conv_chain_code}
  \end{subfigure}
  \hfill
  \begin{subfigure}{0.55\linewidth}
  \centering
  \includegraphics[width=0.8\linewidth]{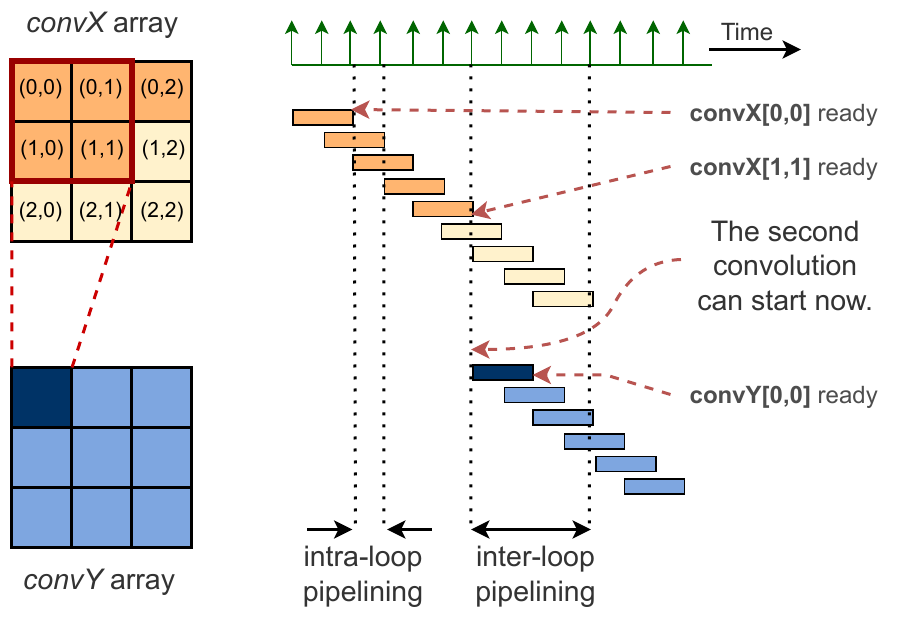}
  \caption{Timing diagram of the pipelined execution.}
  \label{fig:conv_chain_sched}
  \end{subfigure}
  \caption{Example of a producer and consumer loop nest. The consumer loop can
  start execution when the producer has written the first two pixels in row R1
of array convX.}
  \label{fig:conv_chain}
\end{figure*}

In this paper, we introduce a novel ILP-based static scheduler for high-level
synthesis that facilitates pipelining of loop iterations, as well as pipelining
across loop nests with producer-consumer relationships, even in the presence of
complex affine data access patterns between the producer and consumer loop
nests. Furthermore, our scheduler is capable of handling scenarios with multiple
producers and consumers. It is specifically designed for affine workloads with
constant loop bounds.

We implement our scheduler in the HIR framework~\cite{kingshuk21hir}, which is
built on top of the CIRCT project~\cite{circt}. CIRCT is an LLVM incubator
project that utilizes the MLIR infrastructure~\cite{mlir21cgo,mlir2020arxiv} for
developing compilers for high-level synthesis. To lower HLS kernels from C to
MLIR, we extend the Polygeist frontend~\cite{moses21polygeist} with HLS-specific
pragmas.
We make the following key contributions:
\begin{itemize}
  \item We build an ILP-based automatic scheduler that can perform pipelining
  across producer-consumer loops in the presence of arbitrary affine memory
  access patterns.
  \item We integrate our scheduler with the HIR~\cite{kingshuk21hir} compiler
  framework, and extend the Polygeist frontend with HLS-related pragmas to
  enable an end-to-end C-to-SystemVerilog compilation flow.
  \item We quantify the performance and resource usage of our ILP-based
  scheduler on a set of benchmarks. Our evaluation demonstrates that our
  compiler can outperform Vitis HLS (a state-of-the-art commercial HLS compiler)
  while consuming fewer hardware resources.
\end{itemize}

The rest of this paper is organized as follows. In Section~\ref{sec:motivation},
we motivate our work and contrast our approach against
runtime-synchronization-based dataflow optimization techniques.
Section~\ref{sec:compiler} explains the overall
architecture of the MLIR-based compiler. In Section~\ref{sec:ilp}, we discuss
the ILP formulation used by our scheduler.  Experimental results are reported
in Section~\ref{sec:evaluation}. We discuss the related work in
Section~\ref{sec:related-work} and conclude in Section~\ref{sec:conclusions}

\section{Motivation}
\label{sec:motivation}
There are many image-processing and machine-learning workloads that
can benefit from FPGA-based custom accelerators. Often these workloads are
composed of multiple smaller kernels such as convolution and matrix
multiplication.
These kernels may have a producer-consumer relationship among them.
Figure~\ref{fig:conv_chain_code} shows an example of two convolution kernels
applied on an image successively. The second convolution is dependent on the
output of the first convolution. The Unsharp mask algorithm contains this type
of convolution chaining. Machine learning models also often chain multiple
tensor-level operations together. 

Figure~\ref{fig:conv_chain_code} shows an example of producer-consumer loop nests. 
The \textit{conv\_xy} function performs two consecutive convolutions on the input
image.  The output of the first convolution is written to the \textit{convX}
array by the first loop-nest. The second loop-nest consumes the \textit{convX}
array as input for the second convolution.
To improve performance, we unroll the inner two loops and pipeline
the \textit{j-loop} in both convolutions.
Figure~\ref{fig:conv_chain_sched} highlights the elements of the array
\textit{convX} that are required to calculate the first element of the
\textit{convY} array. It also shows the timing diagram of iterations of the
\textit{j-loop}. Each iteration of the j-loop is independent of the previous iterations.
Thus, before an iteration of the loop completes, the next iteration can be
started as shown in the timing diagram. This leads to overlapped execution
of multiple \textit{j-loop} iterations, improving parallelism.
In addition to pipelining between loop iterations (\textit{intra-loop pipelining}), there
is also an opportunity to pipeline between the producer and consumer loop nests
(\textit{inter-loop pipelining}).
The second convolution (consumer loop nest) does not have to wait for the first
convolution to complete. It can start as soon as enough data is available to
start the calculation. Pipelining between the producer and consumer
convolutions leads to overlapped execution as shown in
Figure~\ref{fig:conv_chain_sched}, which further improves parallelism.

Vitis HLS (also known as Vivado HLS) is a state-of-the-art high-level synthesis
compiler developed by Xilinx for their FPGAs.
Vitis HLS offers support for both intra-loop pipelining, as well as pipelining
across producer-consumer loops. 
However, the dataflow optimization feature of Vitis HLS for pipelining the
execution of producer-consumer loops within a single function invocation has
certain limitations:
\begin{itemize}
  \item The consumer must read the data in the same order in which the producer
    generates it.
  \item \textbf{Single-producer-single-consumer (SPSC)} \quad The intermediate
    arrays used to communicate between the producer and consumer tasks must
    have only one producer (i.e., only one task can write to it) and one
    consumer (only one task can read from it).
  \item The intermediate arrays must be instantiated in the function itself.
    Array arguments to a function can not be used to write the intermediate
    values.
\end{itemize}

In Figure~\ref{fig:conv_chain_code}, the single-producer-single-consumer (SPSC)
requirement is satisfied since the \textit{intermediate array}, \textit{convX},
is produced by the first convolution and consumed by the second. However, the
second convolution does not read the \textit{convX} array in the same order in
which the first convolution writes to it. 
In fact, since the consumer task is a stencil operation, the number of reads on
\textit{convX} is more than the number of writes.  Due to the abundance of
stencil operations in both image processing and machine learning applications,
this is a very common scenario.

To solve this problem, we need to first understand why the Vitis HLS
\textit{dataflow} optimization has the above-mentioned limitations. For each
producer-consumer task pair, Vitis HLS dataflow optimization checks the memory
accesses patterns of the intermediate array. If the elements of the intermediate
array are read by the consumer in the order in which they are written by the
producer, then Vitis HLS can replace the array with a FIFO without altering
the program's behavior. The FIFO ensures that the consumer gets the data in
the same order in which the producer is generating it. Now, Vitis HLS can start
the execution of the consumer loop along with the producer loop without having
to worry about data dependence violation. 
If the producer has not generated and pushed the next value to the FIFO the
data, the consumer automatically stalls while attempting to read that value from
the empty FIFO.
Thus the read-after-write dependence between the producer and consumer tasks in
the original program is enforced at runtime by the FIFO's synchronization logic.
If, in the original program, the order of reads and writes to the intermediate
buffer do not match, then Vitis HLS can not replace the intermediate array with
a FIFO. 
In such cases, it instantiates a ping-pong buffer to replace the intermediate
array. This optimization helps in pipelining between multiple invocations of the
function but does not pipeline between the producer and consumer tasks within a
single function invocation. Thus, in absence of the same read and write order,
the dataflow optimization does not improve the overall performance/latency of a
single function invocation.

When multiple consumer loops read from the same intermediate array, each
consumer loop has to wait unless all other consumers have read the current data
before reading the next value. As a result, using a single FIFO with multiple
consumers may lead to unexpected stalls, and even deadlock in the presence of
additional data dependencies between the consumers. Vitis HLS does not duplicate
the FIFOs for multiple consumers. This is probably to ensure that the BRAM usage
does not explode after the dataflow optimization. The FIFO-based implementation
can not handle multiple producers either since at runtime, two producers running
in parallel will insert the data in arbitrary order, and the consumer will not
be able to read the data in the expected order.  Due to the above-mentioned
limitations of the FIFO-based approach, \textit{dataflow} optimization in Vitis
HLS is limited to single-producer-single-consumer (SPSC) workloads.

Similarly, if the intermediate array was accessed via function argument, the
optimizer would not have access to the array and would not be able to replace it
with a FIFO, leading to the third constraint. This leads to the two hard
constraints for dataflow optimization. 

Based on the above discussion, we can identify two key attributes of the
dataflow optimization,
\begin{itemize}
  \item Vitis HLS performs a very basic static analysis just to check if the
    producer and consumer access the intermediate array in the same order. 
    Its static analysis does not handle more complex data access patterns.
  \item The dataflow optimization relies on runtime synchronization to enforce
    read-after-write dependencies between producer and consumer tasks.
\end{itemize}

While the lack of better static analysis leads to missed parallelization
opportunities for Vitis HLS in the presence of complex memory access patterns,
enforcing memory dependence using runtime synchronization can lead to more
resource usage. We will further quantify the effect of these design decisions on the
performance and resource usage in Section~\ref{sec:results}. 

In this paper, we propose a novel approach to scheduling that focuses on
workloads with affine memory accesses and constant loop bounds. By limiting the
scope of our problem to these constraints, we are able to statically schedule
the synthesized hardware design. Confining memory accesses to affine functions
of loop induction variables allows for precise memory dependence analysis. We
argue that kernels in many relevant domains, such as image processing and
machine learning, exhibit affine memory accesses. Furthermore, kernels optimized
for high-level synthesis often have constant loop bounds. For example, a matrix
multiplication implemented as a systolic array would require constant loop
bounds as the systolic array cannot have variable dimensions. In cases where
loop bounds are unknown, partial unrolling may require an additional cleanup
circuit to handle the last few iterations. A common practice is to tile the
workload and sequentially run each tile of fixed size on the accelerator,
exploiting parallelism within a tile to improve performance.

The advantage of our approach is that we can pipeline producer-consumer loop
nests even if the design does not follow the single-producer-single-consumer
(SPSC) rule.  Furthermore, our scheduler can handle more complex memory access
patterns such as in the chain of convolution in
Figure~\ref{fig:conv_chain_code} to generate overlapped schedules like in
Figure~\ref{fig:conv_chain_sched}. 

\section{The compiler pipeline}
\label{sec:compiler}

\begin{figure}
  \centering
  \includegraphics[width=0.65\linewidth]{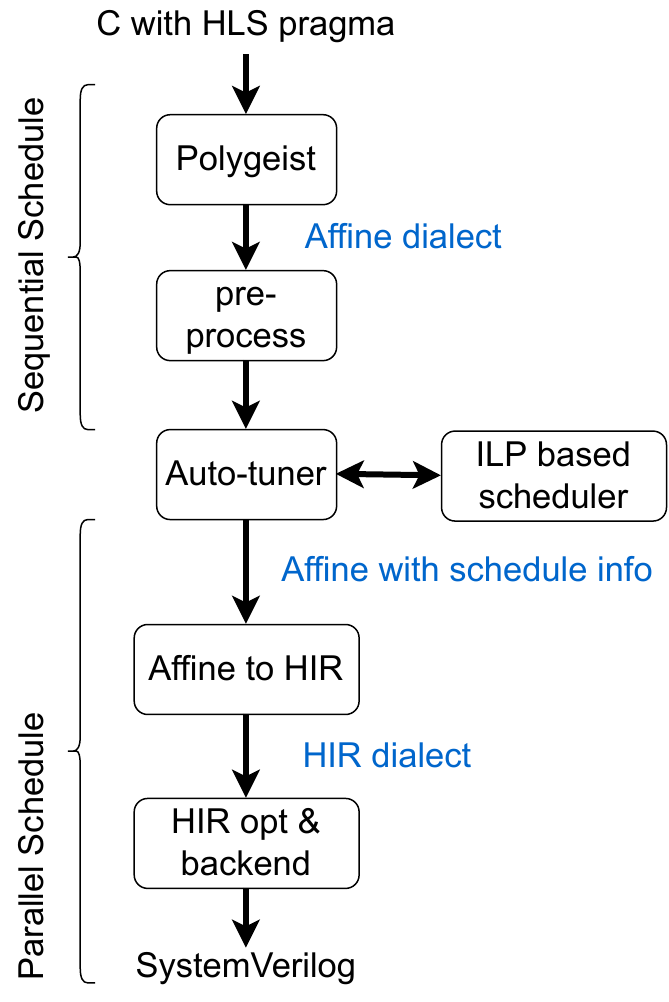}
  \caption{MLIR-based HLS compiler.}
  \label{fig:hir_compiler}
\end{figure}

Figure~\ref{fig:hir_compiler} depicts the comprehensive compiler pipeline based
on MLIR. We utilize the Polygeist~\cite{moses21polygeist} C/C++ frontend to
lower C programs to the affine dialect.  The programmer specifies the target
pipelining and hardware resources using pragmas in the C program. In case the
programmer does not provide explicit specifications, we have implemented a
basic auto-tuner to search for different loop pipelining options. Additionally,
we have implemented an affine-to-HIR pass to lower the program to the HIR
dialect. The HIR backend is then used to optimize the design, such as bit-width
optimization, and generate SystemVerilog. Figure~\ref{code:conv} illustrates the
input and output to the scheduling and affine-to-HIR pass for a convolution
kernel.

\subsection{Polygeist frontend}
We utilize the \textit{C} language as the input programming language for our
compiler toolchain. The Polygeist~\cite{moses21polygeist} frontend is employed
to lower \textit{C} programs into MLIR. We have extended the Polygeist compiler
to include support for HLS-related pragmas, akin to those found in Vitis HLS. We
support the following pragmas in our compiler:

\begin{itemize}
  \item {\bf pipeline} pragma is used to specify the loop initiation interval.
  \item {\bf unroll} pragma is used to specify whether a loop is unrolled or
    not. We only support complete unrolling at present. Partial unrolling can
    be achieved by manually strip-mining the loop and unrolling the resulting
    inner loop completely.
  \item {\bf bind\_storage} pragma is used to specify the type of hardware
    buffer such as block RAM or LUT used to implement an array and the number
    of ports.
  \item {\bf array\_partition} pragma is used to specify memory banking along a
    specific dimension of a multidimensional array. We only support complete
    partitioning of a dimension. Block and cyclic partitioning are not
    supported yet.
  \item {\bf interface} pragma defines the type of interface used to implement
    a function argument. We use this pragma with array arguments to specify the
    number of memory ports and latency of the read/write operation.
  \item {\bf bind\_op} pragma is used to specify the name of the external
    Verilog module which implements a specific operation. For instance,
    floating point operations use Xilinx's Floating point IP. We instantiate
    the IP in a wrapper Verilog module and then use this pragma to specify the
    name of the module and the latency of the floating point operation. This
    allows us to use external FPGA vendor-specific IPs for efficient
    implementation of floating point arithmetic.
  \item {\bf extern\_func} pragma specifies the latency of external functions.
    We can use this pragma to call external Verilog modules. Currently, we only
    support external modules with one output and a constant latency.
\end{itemize}

The Polygeist frontend generates programs in the affine dialect, with pragma
information preserved as attributes.
Following that, a preprocessing pass is applied to convert all floating-point
operations into function calls, utilizing information from the bind\_op pragma.
It also inserts appropriate function declarations into the affine program and
inlines function calls for which definitions are available. The resulting output
of the preprocessing for the one-dimensional convolution is depicted in
Figure~\ref{code:conv_pre}.

Next, we proceed with scheduling, aiming to identify a schedule that satisfies
the programmer-specified initiation intervals (II) without altering the
semantics of the input sequential affine program. The autotuner employs a simple
binary search technique to determine the optimal II for each loop that lacks a
programmer-specified II. For each target II, the scheduler is executed to verify
if the design can be feasibly scheduled. The outcome of this step is a parallel
program.

The affine-to-HIR pass transforms the affine program to HIR using the scheduling
information, and the \textit{binding} pragma attributes which determine the type
of hardware buffers to be used for each multi-dimensional array.
Figure~\ref{code:conv_hir} illustrates the scheduled HIR design for the
one-dimensional convolution. The initiation interval of this design cannot be
reduced below seven clock cycles due to a loop carried dependence between the
\textit{store} and the \textit{load} operations on $\%arg0$. Within the same
iteration of the j-loop, there is a six-cycle gap between a \textit{load} and a
\textit{store} on $\%arg0$, resulting from a one-cycle \textit{load} delay and a
five-cycle delay of the floating point \textit{add} operation. To maintain the
read-after-write dependence between the \textit{load} and \textit{store}
operations, the next iteration's \textit{load} cannot be scheduled before the
current iteration's \textit{store} is completed, which requires an additional
clock cycle. As a result, the next iteration can only commence after a delay of
seven clock cycles. For the sake of brevity, certain operations such as delay
and casting operations are omitted in Figure~\ref{code:conv_hir}.

\begin{figure*}[!htb]
  \centering
    \begin{subfigure}[b]{0.44\linewidth}
\begin{lstlisting}[language=MLIR,breaklines=true,frame=bt]
func.func private @mul_f32(f32, f32) 
  -> (f32 {hir.delay = 4}) 

func.func private @add_f32(f32, f32) 
  -> (f32 {hir.delay = 5}) 

func.func @conv(%arg0: memref<16xf32> 
  {hir.memref.ports = [{rd_latency = 1 : i64, 
                        wr_latency = 1 : i64}]}, 
  %arg1: memref<16xf32>, 
  %arg2: memref<2xf32>){

  affine.for %arg3 = 0 to 16 {
    affine.for %arg4 = 0 to 2 {
      %0 = affine.load %arg0[%arg3] 
      %1 = affine.load %arg1[%arg3 + %arg4] 
      %2 = affine.load %arg2[%arg4] 
      %3 = func.call @mul_f32(%1, %2)  
      %4 = func.call @add_f32(%0, %3)  
      affine.store %4, %arg0[%arg3]
    } {II = 7 : i64}
  } {II = 14 : i64}
  return
}
\end{lstlisting}
      \caption{After preprocessing.}
      \label {code:conv_pre}
    \end{subfigure}
    \begin{subfigure}[b]{0.52\linewidth}
\begin{lstlisting}[language=MLIR,breaklines=true,frame=bt]
hir.func.extern @mul_f32 at %arg2 (%arg0 : i32
  , %arg1 : i32) -> (%out : i32 delay 4)
hir.func @conv at %arg3 (
  %arg0 : !hir.memref<16xi32> 
    ports [{rd_latency = 1, wr_latency = 1}], 
  %arg1: ...,
  %arg2: ...){
  %t_end = hir.for %i = %c0 to %c16 step %c1 
  iter_time( %arg5 = %arg3){
    %res, %t_end_0 = hir.for %j = %c0 to %c2 step %c1 
      iter_time( %arg8 = %arg5){
      %4 = hir.load %arg0[port 0] [%i] at %arg8+4  
      %s = comb.add %i, %j : i4
      %8 = hir.load %arg1[port 0] [%s] at %arg8  
      %10 = hir.load %arg2[port 0] [%j] at %arg8  
      %12 = hir.call @mul_f32(%8, %10) at %arg8+1
      %sum= hir.call @add_f32(%4, %12) at %arg8+5
      hir.store %sum to %arg0[port 0] [%i] 
                              at %arg8 + 10 
      hir.next_iter at %arg8 + 7 : (i64)
    }
    hir.next_iter at %arg5+14
  }
  hir.return
} 
\end{lstlisting}
      \caption{After scheduling and lowering to HIR.}
      \label {code:conv_hir}
    \end{subfigure}
      \caption{Lowering of convolution from C to HIR.}
      \label {code:conv}
\end{figure*}

\subsection{HIR backend}
The HIR dialect~\cite{kingshuk21hir} explicitly captures the schedule of the
design. The HIR compiler backend automatically generates the required state
machines in Verilog to implement the schedule. This makes HIR a good target
after our scheduling pass. We can directly encode the ILP-generated parallel
schedule using HIR's time variables.

Figure~\ref{code:conv_hir} shows an example of the HIR dialect. The key feature
of the HIR dialect is the time variable. Each operation is associated with a
start time. Each region is associated with a time variable. The time variable of
a region denotes the time at which the region is scheduled to start execution.
Operations within the region are scheduled relative to this time variable. For
example, the time variable $\%arg8$ within the j-loop body denotes the time at
which the current iteration of the loop starts.

\section{ILP formulation}
\label{sec:ilp}
In this section, we outline our ILP formulation for the scheduling problem in
our compiler toolchain. The scheduler is responsible for transforming programs
from the affine dialect to the HIR dialect. The original program in the affine
dialect has a sequential schedule, but our goal is to create a parallel schedule
during the transformation to the HIR dialect in order to effectively utilize
hardware resources. The target initiation intervals for each loop are specified
using loop attributes in the affine dialect.

The scheduler's objective is to determine the start time of each operation in
the generated HIR dialect such that the target initiation intervals are
achieved, loop nests are maximally overlapped for improved parallelism, and
memory dependencies are not violated to ensure that the semantics of the 
original sequential program are preserved after automatic parallelization.

To tackle this problem, we represent the scheduling task as a set of ILP
formulations. We first solve smaller ILPs, known as \textit{memory-dependence
ILPs}, for each conflicting memory access (i.e., load/store operations on the
same array).  The solutions of these \textit{memory-dependence ILPs} are then
used to formulate the \textit{scheduling ILP}. The \textit{scheduling ILP}
calculates the start time of each operation in a way that ensures data
dependencies are not violated.

An operation in the program is represented by $S*$ (for example, $S0$). 
An operation may be executed multiple times. 
Each execution of an operation $S$ is referred to as a \textit{dynamic instance}
of the operation, denoted by $S(i,j,k)$ where $i,j,k$ represent the values of
the loop induction variables of the enclosing loops.
This uniquely identifies a dynamic instance of the operation. We denote the
initiation interval specified for a loop with induction variable $i$ as $II_i$.
The start time of each operation is specified relative to its parent region. So
the start time of operation $S0$ inside a loop represents the delay after which
$S0$ will execute relative to the start time of the current loop iteration and
is denoted by $t_{S0}$. Loops are also treated like any other operation. The
start time of a loop with induction variable $i$ is denoted by $t_i$. These
variables denoting the start time of an operation are called time variables.  We
now explain our ILP formulations for \textit{intra-loop} (overlapping iterations
of a loop) and \textit{inter-loop} (overlapping execution of multiple loops)
pipelining with two examples.

\subsection{Intra-loop dependence}
\begin{figure}[h]
\begin{lstlisting}[language=C++,breaklines=true,frame=bt,basicstyle=\footnotesize]
// floating point addition delay=5
// floating point multiplication delay=4
// Load and store to A,B,C take one cycle each.

// Loop-i starts first iteration at time T_i
for (int i = 0; i < 10; i++) {
  // Initiation interval = II_i.
  for (int j = 0; j < 10; j++) {
    // Initiation interval = II_j.
    for (int k = 0; k < 10; k++) {
      // Initiation interval of loop i is II_k.
      a = load(A[i][j]);
      b = load(B[i][j]);
      m = a * b;

    // S0 is executed t_s0 cycles
    // after the current k-loop starts.
    S0:
      c_prev = load(C[i][j]);
    // S1 is executed t_s1 cycles after
    // the current k-loop starts.
    S1:
      c = c_prev + m;
    // S2 is executed t_s2 cycles after
    // the current k-loop starts.
    S2:
      store(c, C[i][j]);
    }
  }
}
\end{lstlisting}
  \caption{Intra-loop memory dependence.}
  \label {code:intra_loop_dep}
\end{figure}

Figure~\ref{code:intra_loop_dep} shows a matrix multiplication kernel. 
We note that there is a write-after-read (WAR) dependence from statement S0 to
statement S2 and a read-after-write (RAW) dependence from S2, to S0 of next
iteration via the \textit{C} array. More precisely, we have a dependence from
$S2(i',j',k')$ to $S0(i,j,k)$ iff,

\begin{itemize}
  \item{\bf Address conflict:} Both statements are accessing same element of array $C$,
    \begin{equation}
      i=i', j=j'. 
      \label{eq:addr_conflict}
    \end{equation}
  \item{\bf Happens before:} If the program ran sequentially, $S2(i',j',k')$
  would occur before $S0(i,j,k)$,
    \begin{equation}
      i'*100+j'*10+k' > i*100+j*10+k.
      \label{eq:happens_before}
    \end{equation}
\end{itemize}

We want to ensure that after scheduling, $S2(i',j',k')$ still occurs after
$S0(i,j,k)$.

Given the target initiation intervals of each loop, the time instant at which
the statement \textit{S0(i,j,k)} will be executed post-scheduling, is given by:

\begin{equation}
  T_{S0}(i,j,k) = t_i+i*II_i+j*II_j+k*II_k+t_{S0}.
  \label{eq:sched_time}
\end{equation}

For a valid schedule, we need to ensure that:
\begin{equation}
  T_{S0}(i',j',k') \geq T_{S2}(i,j,k)+1. 
  \label{eq:valid_sched}
\end{equation}

The extra one cycle is because the store operation on array \textit{C} due to statement
$S2$ would take one cycle to complete. 
At this point, we may think that in Equations~\ref{eq:addr_conflict},
\ref{eq:happens_before} and \ref{eq:valid_sched}, we have the necessary set of
linear constraints to ensure that the resulting ILP solution does not violate
the RAW dependence from $S2$ to $S0$. But this is not true. If we put
these equations in our ILP formulation, the ILP solver would find one
possible set of values for $(i,j,k,i',j',k')$ such that the constraints are
satisfied. But for a valid schedule, we need that every $(i,j,k,i',j',k')$ that
satisfies Equation~\ref{eq:addr_conflict} and \ref{eq:happens_before}, also
satisfy Equation~\ref{eq:valid_sched}. To formulate the correct ILP, we expand
Equation~\ref{eq:valid_sched} using Equation~\ref{eq:sched_time} and reorder the
terms as follows:

\begin{eqnarray}
  t_{S2} - t_{S0} & \leq & (i'*II_i+j'*II_j+k'*II_k) \nonumber \\ 
  &&- (i*II_i + j*II_j +k*II_k) - 1.
  \label{eq:valid_sched2}
\end{eqnarray}

We define a new variable called \textit{slack} as follows:
\begin{equation} 
  \begin{split}
    slack = minimize(&(i'*II_i+j'*II_j+k'*II_k) \\&- (i*II_i + j*II_j +k*II_k) - 1).
  \end{split}
  \label{eq:slack}
\end{equation}

such that $(i,j,k,i',j',k')$ satisfy Equation~\ref{eq:addr_conflict} and \ref{eq:happens_before}.

If $t_{S2} - t_{S0} \leq slack$ then the dependence from $S2$ to $S0$ is never violated. We
calculate $slack$ by solving Equation~\ref{eq:slack} as a
minimization ILP problem with Equation~\ref{eq:valid_sched2} and
\ref{eq:happens_before} as the constraints. We also add constraints for the
loop bounds on the induction variables $(i,j,k,i',j',k')$.
For each potential memory dependency, we create such an ILP, called the
\textit{memory dependence ILP}. If the ILP does not have a solution then there
is no actual dependency. Otherwise, we use the calculated slack to add
constraints on the time variables in the \textit{scheduling ILP}. For port
conflicts, we assume that all operations on the same port of the same memory bank
have a data dependence. This added dependence ensures that the resulting
schedule does not have port conflicts, i.e., it does not schedule two memory
access operations on the same port of the same memory bank in the same clock
cycle.

In addition to these constraints related to memory dependence, we also add
constraints related to SSA dependence and operator delay in the scheduling ILP.
For example, due to the SSA variable \textit{c\_prev}, there is a dependence
from statement $S0$ to $S1$. Since a load operation takes one clock cycle to
complete, $S1$ must wait for one cycle before $S0$. We capture this using the
constraint: $$t_{S1}-t_{S0}>1.$$

Note that $t_{S0}$ and $t_{S1}$ are the start time of the respective operations
relative to the start time of the current iteration of the k-loop. But since
both the operations are in the same loop-nest, this constraint is enough to
ensure that $S1(i,j,k)$ happens after $S0(i,j,k)$ for all values of $(i,j,k)$.

\subsection{Inter-loop dependence}
\label{sec:inter-loop-dependence}
\begin{figure}[h]
\begin{lstlisting}[language=C++,breaklines=true,frame=bt,basicstyle=\footnotesize]
// Loop-i1 starts at t_i.
for (int i = 0; i < 10; i++)
  // Initiation interval = II_i.
  for (int j = 0; j < 10; j++)
    // Initiation interval = II_j.

    // S1 is scheduled at t_s1 cycles after the 
    // current iteration of j starts. 
    S1: store (val, A[i][j]);


// Loop-i2 starts at t_u
for (int u = 0; u < 10; u++)
  // Initiation interval = II_u.
  for (int v = 0; v < 10; j++)
    // Initiation interval = II_v.
    
    // S2 is scheduled at t_s2 cycles after the 
    // current iteration of v starts. 
    S2: val = load (A[u][v])
\end{lstlisting}
  \caption{Inter-loop memory dependence.}
  \label {code:inter_loop_dep}
\end{figure}

In the previous example, we discussed how we handle memory dependence within a
loop nest. Now we will discuss how our technique generalizes for memory
dependence across loop nests. Figure~\ref{code:inter_loop_dep} shows an example
of a pair of producer-consumer loop nests. Statement $S2$ has a
read-after-write dependence on statement $S1$. Similar to the previous example,
we calculate the absolute time $T_{S1}$ and $T_{S2}$ as follows:

\begin{equation}
  T_{S1}(i,j) = t_i+i*II_i+j*II_j + t_{S1},
\end{equation}

\begin{equation}
  T_{S2}(u,v) = t_u+u*II_u+v*II_v+ t_{S2}.
\end{equation}

For an actual memory dependence between two dynamic instances $S1(i,j)$ and
$S2(u,v)$, they must access the same element of the array A, i.e., $$ u=i,v=j.$$

Unlike the previous example, we do not need the \textit{happens before}
constraint here because all instances of $S1$ happen before all the instances
of $S2$ in the \textit{sequential schedule}. In general, the \textit{happens
before} criterion is required only if there is at least one common loop between
the two statements. 

The memory dependence is not violated iff:

\begin{equation}
  \begin{split}
    &T_{S2}(u,v) \geq T_{S1}(i,j) + 1\\
    &\forall \quad 0\leq u\leq 10 \textrm{,} \quad  0\leq u\leq 10 \\
    & \quad \quad 0\leq i\leq 10 \textrm{,} \quad  0\leq j\leq 10, \\
    & \quad \quad u=i \textrm{,} \quad  v=j.
  \end{split}
\end{equation}

Similar to the prior example, we calculate the slack using the following ILP,
\begin{equation}
  \begin{split}
    slack = min&imize(u*II_u+v*II_v - i*II_i-j*II_j - 1)\\
    &\textrm{such that} \quad 0\leq u\leq 10 \textrm{,} \quad  0\leq u\leq 10 \\
    & \quad \quad 0\leq i\leq 10 \textrm{,} \quad  0\leq j\leq 10, \\
    & \quad \quad u=i \textrm{,} \quad  v=j.
  \end{split}
\end{equation}

The constraint added to the scheduling ILP to ensure that the memory dependence is not violated is:
$$t_i+t_{S1} - t_u - t_{S2} \leq slack.$$

\subsection{Minimization objective}
\begin{figure}[h]
\begin{lstlisting}[language=MLIR,breaklines=true,frame=bt]
//Inefficient schedule.
hir.for %i:i6 = %c0 to %c32 step %c1 
    iter_time(%ti=%t){
  %x = hir.load %A[port 0][%i]
  %xx = hir.delay %x by 999 at %ti
  hir.store %xx to %B[%i] at %ti+1000
  hir.next_iter at %ti+1
}

//Efficient schedule.
hir.for %i:i6 = %c0 to %c32 step %c1 
    iter_time(%ti=%t){
  %x = hir.load %A[port 0][%i]
  hir.store %x to %B[%i] at %ti+1
  hir.next_iter at %ti+1
}
\end{lstlisting}
  \caption{Resource-inefficient and -efficient schedules.}
  \label {code:delay_sched}
\end{figure}

The scheduling ILP guarantees a valid schedule but it does not ensure resource
efficiency. Consider the example in Figure~\ref{code:delay_sched}. In the first
loop, the schedule is valid but there is an unnecessary delay of a thousand cycles
between the load and the dependent store op. To ensure that the value
of variable $x$ is not overwritten by the time the store operation happens, a
delay of 999 cycles is introduced using the \textit{hir.delay} operation. This is
implemented in hardware using shift registers which is a waste of resources. If
the store op is scheduled at $ti+1$, then a delay is not required as the load
op produces the value at $ti+1$. 
In the ILP formulation, we calculate the delay required by an SSA variable as
the difference in the start time of the operation consuming the SSA variable and
the operation producing it. We use the sum of these delays as our minimization
objective to reduce shift-register usage in the final design.

\section{Evaluation}
\label{sec:evaluation}
In this section, we will compare the performance and FPGA resource usage of the
HIR compiler with our ILP-based scheduler against the Vitis HLS (formerly Vivado
HLS) compiler. The objective of our evaluation is to quantify the advantage of
multi-dimensional (inner-loop and producer-consumer) pipelining and resource
usage improvement due to static scheduling.
We have open-sourced our work on github~\cite{pipeflow}.
\subsection{Methodology}
We extend the Polygeist C/C++ frontend with pragmas similar to Vitis
HLS. The Polygeist compiler frontend inserts the pragma information as
attributes in the generated Affine dialect. We use the HIR compiler as our
backend to generate SystemVerilog. Our ILP-based scheduler is invoked when
lowering from the Affine dialect to the HIR dialect. We evaluate the effectiveness
of the resulting end-to-end (C-to-SystemVerilog) compiler on a set of benchmarks
by generating SystemVerilog from both our compiler and Vitis HLS and comparing
their performance and resource usage. For performance evaluation, we use
Vivado's SystemVerilog simulator to calculate the total number of cycles
required for each design to complete execution. We report the resource usage by
running synthesis and implementation using Vivado on the generated SystemVerilog
designs. All generated designs are synthesized for the VC709 Xilinx FPGA board
with a target frequency of 200MHz. The correctness of the generated design is
validated on random inputs by comparing the simulation results with Vitis HLS.
We also verify that our designs can meet the timing closure post-implementation
(i.e. after the place-and-route phase) in Vivado.

\vspace{1em}
\noindent\textbf{Unsharp mask} is an image processing pipeline to sharpen the
image at the boundaries. It involves calculating the image blur along the
\textit{X} and textit{Y} direction and then applying a pointwise sharpening and
masking filter. We use a 32x32 patch in our benchmark.

\vspace{1em}
\noindent\textbf{Harris} is a classic corner detection
algorithm\cite{harris1988acc}. It involves multiple stencil operations such as
calculating gradients along the \textit{x} and \textit{y} axis. We use a 32x32
image patch as input for the benchmark.

\vspace{1em}
\noindent\textbf{DUS} is an image pipeline where we downsample the image by a
factor of two and then upsample it. Downsampling involves blurring and 
upsampling uses linear interpolation, both of which are stencil operations.
Both these operations are performed on each axis separately, resulting in four
loop nests. Downsampling and upsampling are very common in image processing
pipelines and offer a unique challenge since the producer and consumer loop
nests may not have the same number of iterations. We use a 32x32 image patch to
evaluate the benchmark.

\vspace{1em}
\noindent The \textbf{Optical flow} benchmark implements the Lucas-Kanade dense
optical flow algorithm~\cite{lucas1981ijcai}. We implemented the single-scale
version of the algorithm and used a 32x32 image patch for our evaluation. 
The benchmark is a mix of pointwise and stencil operations.

\vspace{1em}
\noindent\textbf{2mm} is a benchmark from the \textit{polybench} suite. It
involves two matrix multiplications in a series. Both the intermediate and the
final matrix are written to the output. This benchmark shows that our compiler
can handle non-stencil memory access patterns as well. We use an 8x8 matrix to
evaluate the benchmark.

\subsection{Results}
\label{sec:results}
In our evaluation, we try to quantify the performance and resource usage of our
compiler and compare it with Xilinx Vitis HLS (formerly Vivado HLS), a
state-of-the-art commercial high-level synthesis compiler.
Specifically, we compare our scheduler's ability to reduce the overall
latency by pipelining across multiple producer-consumer loop nests.

\vspace{1em}
\colorbox{gray!40}{\parbox{0.9\linewidth}{Q. Does overlapped execution of loop
nests provide any meaningful performance improvement?}} \vskip 10pt

\begin{figure}[h]
  \includegraphics[width=0.9\linewidth]{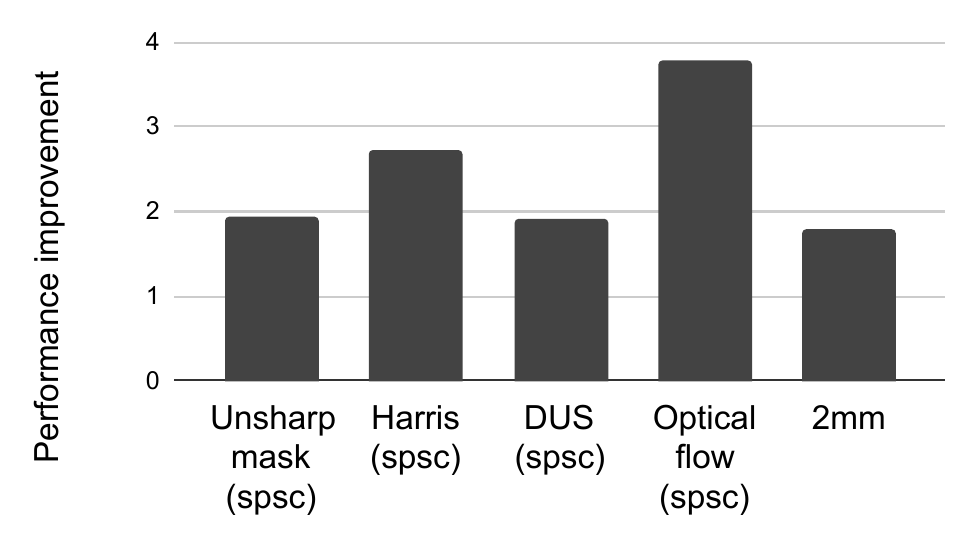}
  \caption{Performance improvement due to overlapped execution of loop nests.}
  \label{fig:perf_overlap}
\end{figure}

A key optimization enabled by our scheduler's multi-dimensional pipelining
is the overlapped execution of producer-consumer loop nests as discussed in
Section~\ref{sec:motivation} and Section~\ref{sec:inter-loop-dependence}. 
But we need to quantify how much performance improvement can be attributed to
this optimization compared to traditional intra-loop pipelining. To understand
this, we generate the scheduled HIR and calculate the total latency (number of
clock cycles) of each loop nest by multiplying the initiation interval of
the outermost loop with its trip count.  We then add these latencies to get the
final latency of the complete kernel. This is the latency of the kernel
if each loop was pipelined but there was no overlapping between different loop
nests. We use this to quantify the performance improvement of a kernel with all
loop nests maximally overlapped (without violating data dependencies) compared
to a kernel with just intra-loop pipelining and no pipelining between loop
nests. As shown in Figure~\ref{fig:perf_overlap}, producer-consumer pipelining
yields between 1.7x and 3.7x performance improvements on top of the intra-loop
pipelining optimization - highlighting the practical importance of
multi-dimensional pipelining.

\vspace{1em}
\colorbox{gray!40}{\parbox{0.9\linewidth}{Q. How does our scheduler compare
against Vitis HLS dataflow optimizations?}}

\begin{figure}[h]
  \includegraphics[width=0.9\linewidth]{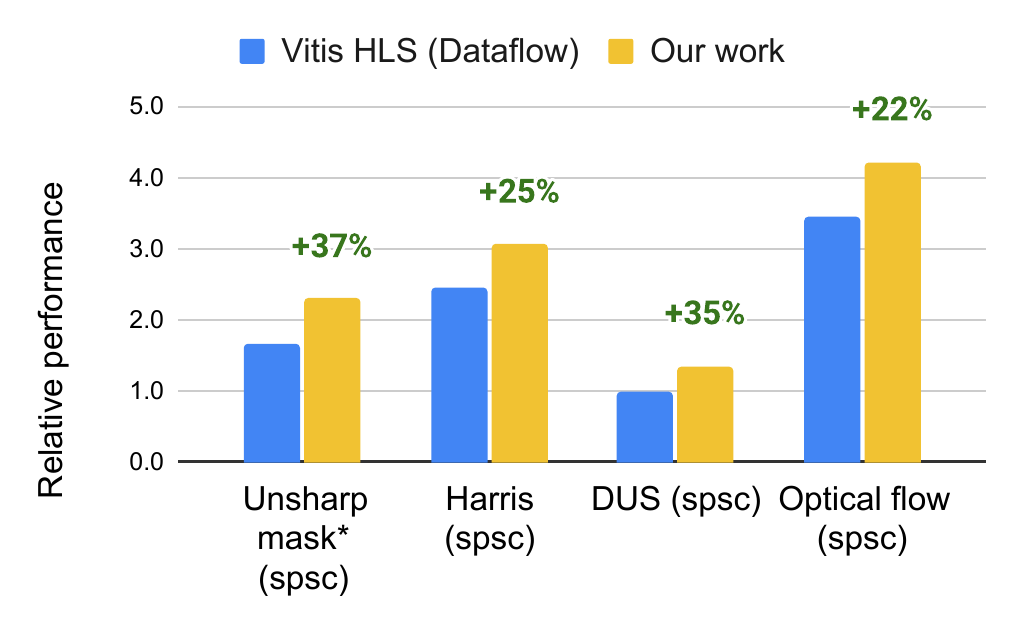}
  \caption{Performance comparison between Vitis HLS (with dataflow directives)
  and our work. The baseline is Vitis HLS without dataflow directives.}
  \label{fig:perf_spsc}
\end{figure}

Vitis HLS's \textit{dataflow} optimization also attempts to overlap the
execution of producer-consumer loop nests. We compare our performance gains
against Vitis HLS in Figure~\ref{fig:perf_spsc}. Vitis HLS \textit{dataflow}
pragma only works if the intermediate arrays between producer-consumer loops
follow the single-producer-single-consumer rule i.e. only one loop nest writes to the
array and only one loop nest reads from it (as discussed in
Section~\ref{sec:motivation}). DUS already satisfies the
single-producer-consumer constraint. We convert \textit{unsharp mask},
\textit{Harris corner detection} and \textit{optical flow} to
single-producer-consumer workloads by inserting copying loops that duplicate the
intermediate arrays that are consumed by multiple loop nests. Yet another
limitation of Vitis HLS dataflow optimization is that it can not handle
read/write to function arguments in the dataflow region.  Due to this limitation
of Vitis HLS, we could not use the 2mm benchmark for this specific evaluation,
since it writes the intermediate matrix multiplication output to one of the
function arguments.

\begin{figure*}
  \centering
  \begin{subfigure}{0.39\linewidth}
    \centering
    \includegraphics[width=\linewidth]{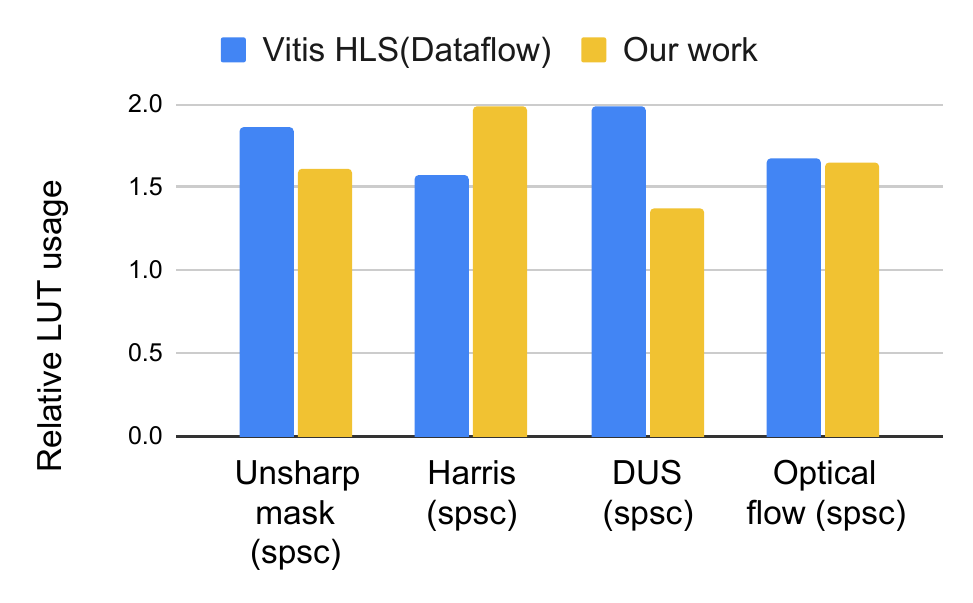}
    \caption{LUT usage.}
    \label{fig:lut_spsc}
  \end{subfigure}
  \begin{subfigure}{0.39\linewidth}
    \centering
    \includegraphics[width=\linewidth]{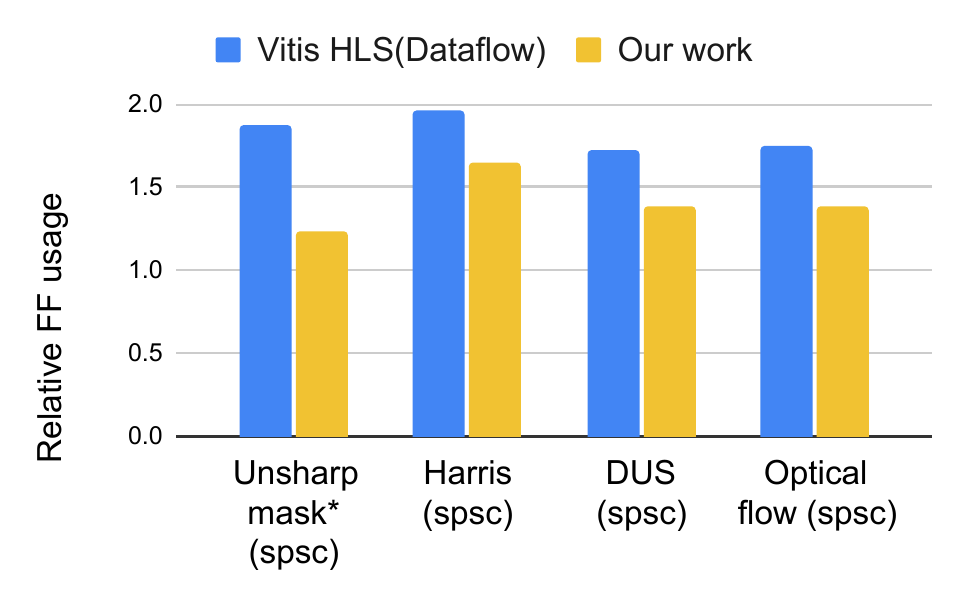}
    \caption{FF usage.}
    \label{fig:ff_spsc}
  \end{subfigure}
  \begin{subfigure}{0.39\linewidth}
    \centering
    \includegraphics[width=\linewidth]{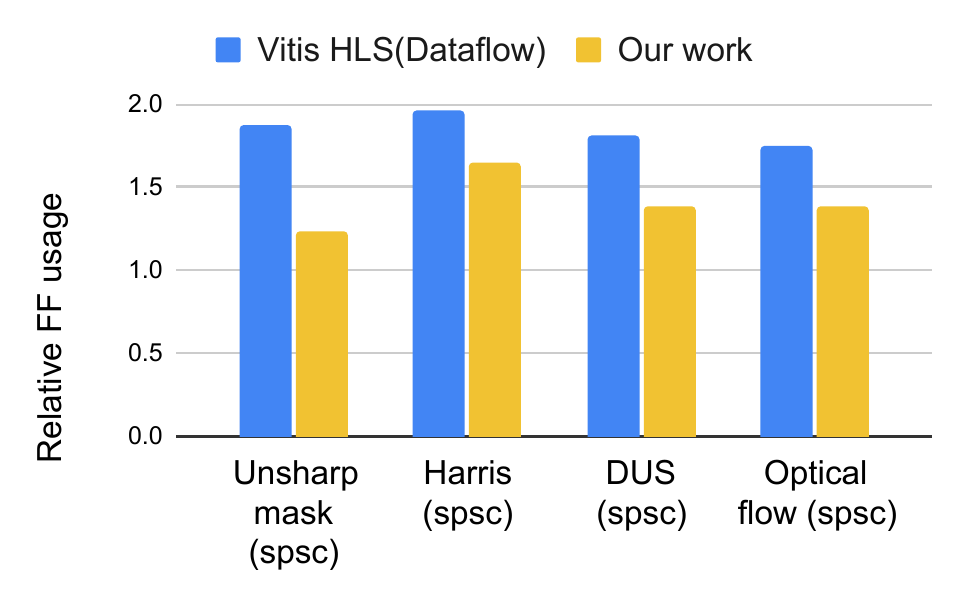}
    \caption{BRAM usage.}
    \label{fig:bram_spsc}
  \end{subfigure}
  \begin{subfigure}{0.39\linewidth}
    \centering
    \includegraphics[width=\linewidth]{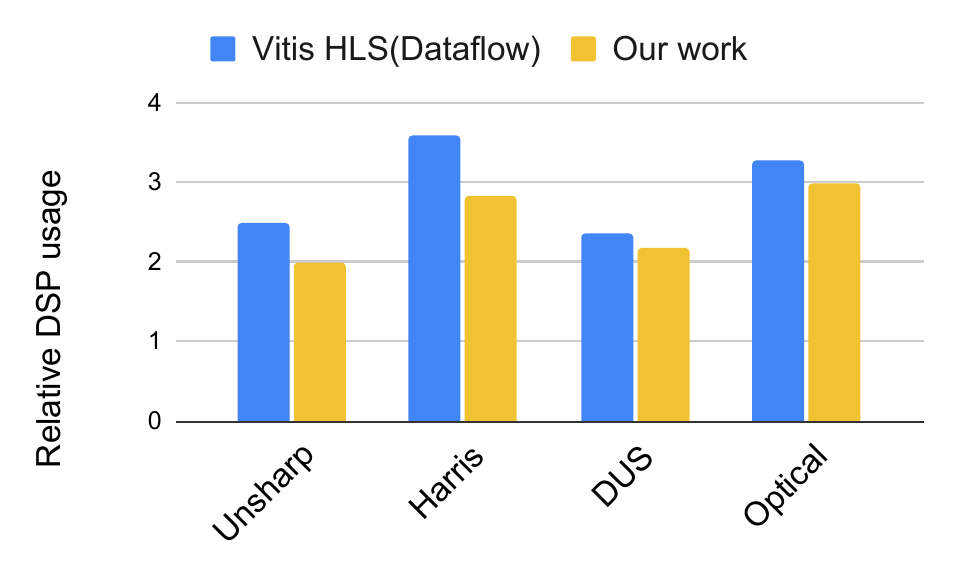}
    \caption{DSP usage.}
    \label{fig:dsp_spsc}
  \end{subfigure}
    \caption{Resource usage of Vitis HLS with dataflow pragma and our work relative to Vitis HLS without the dataflow directives. }
    \label{fig:usage_spsc}
\end{figure*}

Figure~\ref{fig:perf_spsc} reports the performance of both our work and Vitis
HLS with dataflow optimization enabled. The performance numbers are plotted
relative to Vitis HLS without the dataflow optimization directive. We observe
that the Vitis HLS \textit{dataflow} directive indeed improves the overall
latency of the kernels for the single-producer-single-consumer (SPSC) workloads.
However, the designs generated by our compiler provide additional performance
gains of up to $37\%$ on top of the dataflow optimized Vitis HLS designs. This
showcases the effectiveness of our ILP-based scheduler. But the discussion on
performance improvements is incomplete without a detailed comparison of the
hardware resources consumed by the two designs.

\vspace{1em}
\colorbox{gray!40}{\parbox{0.9\linewidth}{
Q. How does our design's resource usage compare to Vitis HLS?}}
\vskip 10pt

Figure~\ref{fig:usage_spsc} shows the resource usage of Vitis HLS dataflow
designs and the designs generated by our compiler relative to Vitis HLS
generated designs without the \textit{dataflow} pragma. As expected, both Vitis
HLS dataflow and our design consumes more resources compared to the non-dataflow
design since they both perform much better than the non-dataflow design. 

Interestingly, our design consumes fewer resources in all benchmarks
(except for LUT usage in Harris corner detection) while, simultaneously,
outperforming the Vitis HLS dataflow optimized design. We see the greatest
improvement in BRAM usage. This can be attributed to the synthesis of ping-pong
buffers by Vitis HLS for more complex dataflow patterns which result in more
BRAM usage compared to the baseline non-dataflow Vitis HLS design. Our design,
on the other hand, does not instantiate any extra RAMs. 

Depending on the data access pattern, Vitis HLS synthesizes either ping-pong
buffers or FIFOs to replace the intermediate arrays accessed by producer and
consumer loops. Runtime synchronization ensures that the consumer reads data
only after the producer has written the data to the intermediate buffer. Though
it is impossible to isolate the exact cause of the extra LUT and FF overhead,
one possible source of the extra resource usage can be attributed to the extra
logic required to implement these synchronizations. On the other hand, our
compiler statically schedules the producer and consumer loops,
avoiding the need for any extra synchronization logic at runtime. 

\vspace{1em}
\colorbox{gray!40}{\parbox{0.9\linewidth}{
Q. What are the limitations of Vitis HLS dataflow optimization and why does our
technique work in those scenarios?
}}
\vskip 10pt

As shown in Figure~\ref{fig:perf_spsc}, Vitis HLS with \textit{dataflow}
optimization does not offer any improvement in the overall latency for
the DUS benchmark. 
This is consistent with our discussion in Section~\ref{sec:motivation} on the
limitations of the \textit{dataflow} optimization pass of Vitis HLS.
Since both upsampling and downsampling loops in DUS read a window of pixels, they
violate the \textit{same-read-write-order} constraint (similar to the
chain-of-convolutions kernel in Figure~\ref{fig:conv_chain_code}).
Because of this, Vitis HLS can not overlap the execution of the loop nests
present in downsampling and upsampling. 
In contrast, our compiler uses an ILP-based scheduler to find the dependence
distances between producer and consumer loop nests and can handle arbitrary
affine accesses on the intermediate arrays. 
As a result, we see a $35\%$ performance improvement in DUS over Vitis HLS. 

Other benchmarks have a mix of loop nests, some of which can and can not be
overlapped by Vitis HLS, resulting in different levels of improvement over the
baseline. For instance, in the unsharp mask benchmark, there are two consecutive
convolution operations that Vitis HLS can not overlap (since the consumer
convolution reads a window of pixels similar to DUS) but it can overlap the
sharpening and masking loop nests which are both pointwise access. Our approach,
on the other hand, can safely overlap/pipeline both types of producer-consumer
pairs. This accounts for the consistently higher performance of the designs
generated by our compiler in all the benchmarks compared to Vitis HLS with
\textit{dataflow} optimizations.

\vspace{1em}
\colorbox{gray!40}{\parbox{0.9\linewidth}{
Q. Can our ILP-based scheduler optimize non-SPSC workloads?
}}
\vskip 10pt

\begin{figure}[h]
  \includegraphics[width=\linewidth]{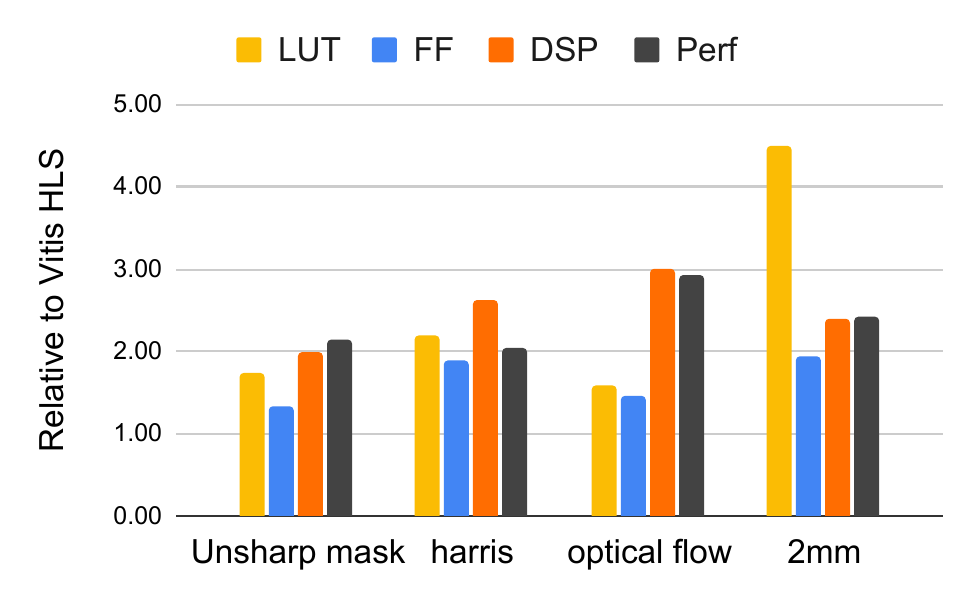}
  \caption{Resource usage and performance of our work relative to Vitis HLS for
  workloads without SPSC dataflow pattern.}
  \label{fig:hir_nospsc}
\end{figure}

We converted our benchmarks to SPSC (single-producer-single-consumer) to enable
dataflow optimizations in Vitis HLS. However, considering that our scheduler
claims to handle non-SPSC workloads, it is natural to question whether we
would observe any performance improvement on the unmodified programs, i.e.
programs with multiple consumers and different data access ordering between the
producer and consumer. Figure~\ref{fig:hir_nospsc} illustrates the effectiveness
of our technique on the unmodified benchmarks, including the 2mm benchmark, as
our compiler can handle function argument accesses as well. We achieved a
speedup ranging from 2x to 2.9x over Vitis HLS without the \textit{dataflow}
pragma (since the \textit{dataflow} optimization in Vitis HLS can not handle
the unmodified non-SPSC workloads).  Vitis HLS can reuse resources, such as DSP units,
across loop nests as they execute sequentially. However, we consume more
resources to achieve additional performance gains.

\section{Related work}
\label{sec:related-work}
In this section, we discuss the previous work on hardware design and high-level synthesis.
\\\\
\textbf{HDLs}
Hardware description languages (HDLs) such as VHDL and Verilog provide
developers with the capability to represent complex designs with fine-grained
control over the instantiated hardware and scheduling. Examples of HDLs
implemented as embedded domain-specific languages (DSLs) on top of existing
software programming languages include Chisel~\cite{dac12chisel},
MyHDL~\cite{myhdl}, and VeriScala~\cite{fpga17veriscala}. These DSLs leverage
the meta-programming capabilities of the host language to represent highly
parameterized hardware designs.
Another approach is taken by BlueSpec SystemVerilog~\cite{memocode04bluespec},
which represents hardware designs as a set of guarded atomic rules, enabling
improved static analysis and code reuse.  
\\\\
\textbf{HLS}
Several HLS compilers, such as Vivado HLS~\cite{vivado}, Intel OpenCL
stack~\cite{intelopencl}, and the LegUp compiler~\cite{fpga11legup}, have repurposed
existing general-purpose software languages like C/C++ and OpenCL for hardware
description, while others, such as Aetherling~\cite{durst2020aetherling}, Dahlia~\cite{pldi20dahlia},
Spatial~\cite{pldi18spatial}, and HeteroCL~\cite{fpga19heterocl}, have opted for
a clean-slate DSL design specialized for high-level synthesis.
Dahlia~\cite{pldi20dahlia} emphasizes performance predictability by using
time-sensitive affine types to check and prevent conflicting resource usage
during compilation. 
Implemented as a DSL embedded in Haskell, Aetherling~\cite{durst2020aetherling} 
compiles data-parallel programs into statically scheduled, streaming hardware
circuits. It proposes sequence data types to encode throughput by specifying
when sequence elements are produced and consumed.
Halide~\cite{pldi13halide} and Polymage~\cite{asplos15polymage} DSLs offer HLS
backends~\cite{fpga20heterohalide,pact16polymagefpga} for FPGAs.
Darkroom~\cite{hegarty2014darkroom} and Rigel~\cite{hegarty2016rigel} implement
a domain-specific high-level synthesis compiler for image processing.
SODA~\cite{iccad18soda}, StencilFlow{cgo21stencilflow} and SASA \cite{tian2023sasa}
use custom architectures and optimizations for stencil computations. 
\\\\
\textbf{Polyhedral optimizations}
The quality of hardware generated by HLS compilers depends heavily on the input
program.  Many polyhedral-based optimization techniques have been proposed in
the past~\cite{asap10alias,fpga13pouchet,fpga13zuo,codes13zuo} that employ
polyhedral techniques to improve the input program for the HLS tool.
These pre-processing loop optimization techniques complement our work.
For instance, \textit{loop interchange} may enable greater overlap between
producer and consumer loops.
AutoSA~\cite{wang2021autosa} and PolySA~\cite{cong2018polysa} utilize polyhedral
optimization to synthesize efficient systolic-array-based hardware designs.
The optimizations in both these works are tailored for kernels that can be
represented as systolic arrays in hardware. In contrast, our work focuses on
pipelining arbitrary affine kernels with constant loop bounds and is not limited
to generating systolic array-based architectures.
\\

Many of the above
DSLs~\cite{iccad18soda,wang2021autosa,fpga20heterohalide,pact16polymagefpga,fpga19heterocl}
and optimization frameworks~\cite{fpga13pouchet,fpga13zuo} use the Vivado HLS
(currently known as the Vitis HLS) compiler, justifying our comparison against
Vitis HLS in the evaluation. Our work is complimentary to these DSLs and optimizations.
We choose the HIR compiler due to the ease of representing static schedules in
HIR intermediate representation, but our ILP-based scheduling technique can be
incorporated in any HLS compiler as another optimization pass.
\section{Conclusions} 
\label{sec:conclusions} 

In this paper, we presented an ILP-based automatic scheduler for the high-level
synthesis of hardware accelerators and integrated it into
HIR~\cite{kingshuk21hir}, an MLIR-based HLS compiler framework. 
Experimental results showed that the multi-dimensional pipelining performed by
our scheduler results in a performance improvement of 1.30$\times$ over
Vitis HLS with \textit{dataflow} optimizations on a representative set of
benchmarks, while consuming fewer resources.
This performance improvement results from our scheduler's ability to overlap
execution of producer-consumer loop nests in the presence of arbitrary affine
access and multiple-producer-multiple-consumer dataflow.

\bibliography{references}

\end{document}